\documentclass{article}
\usepackage{amsmath, amssymb, amsfonts}
\title{Conformal time dependent Painleve-Gullstrand spacetime } 
\author{Hristu Culetu, \\Ovidius University, Dept.of Physics, \\B-dul Mamaia 124, 900527 Constanta, Romania, \\e-mail : hculetu@yahoo.com}

\begin{document}
\numberwithin{equation}{section}
\pagenumbering{arabic}
\maketitle
\newcommand{\fv}{\boldsymbol{f}}
\newcommand{\tv}{\boldsymbol{t}}
\newcommand{\gv}{\boldsymbol{g}}
\newcommand{\OV}{\boldsymbol{O}}
\newcommand{\wv}{\boldsymbol{w}}
\newcommand{\WV}{\boldsymbol{W}}
\newcommand{\NV}{\boldsymbol{N}}
\newcommand{\hv}{\boldsymbol{h}}
\newcommand{\yv}{\boldsymbol{y}}
\newcommand{\RE}{\textrm{Re}}
\newcommand{\IM}{\textrm{Im}}
\newcommand{\rot}{\textrm{rot}}
\newcommand{\dv}{\boldsymbol{d}}
\newcommand{\grad}{\textrm{grad}}
\newcommand{\Tr}{\textrm{Tr}}
\newcommand{\ua}{\uparrow}
\newcommand{\da}{\downarrow}
\newcommand{\ct}{\textrm{const}}
\newcommand{\xv}{\boldsymbol{x}}
\newcommand{\mv}{\boldsymbol{m}}
\newcommand{\rv}{\boldsymbol{r}}
\newcommand{\kv}{\boldsymbol{k}}
\newcommand{\VE}{\boldsymbol{V}}
\newcommand{\sv}{\boldsymbol{s}}
\newcommand{\RV}{\boldsymbol{R}}
\newcommand{\pv}{\boldsymbol{p}}
\newcommand{\PV}{\boldsymbol{P}}
\newcommand{\EV}{\boldsymbol{E}}
\newcommand{\DV}{\boldsymbol{D}}
\newcommand{\BV}{\boldsymbol{B}}
\newcommand{\HV}{\boldsymbol{H}}
\newcommand{\MV}{\boldsymbol{M}}
\newcommand{\be}{\begin{equation}}
\newcommand{\ee}{\end{equation}}
\newcommand{\ba}{\begin{eqnarray}}
\newcommand{\ea}{\end{eqnarray}}
\newcommand{\bq}{\begin{eqnarray*}}
\newcommand{\eq}{\end{eqnarray*}}
\newcommand{\pa}{\partial}
\newcommand{\f}{\frac}
\newcommand{\FV}{\boldsymbol{F}}
\newcommand{\ve}{\boldsymbol{v}}
\newcommand{\AV}{\boldsymbol{A}}
\newcommand{\jv}{\boldsymbol{j}}
\newcommand{\LV}{\boldsymbol{L}}
\newcommand{\SV}{\boldsymbol{S}}
\newcommand{\av}{\boldsymbol{a}}
\newcommand{\qv}{\boldsymbol{q}}
\newcommand{\QV}{\boldsymbol{Q}}
\newcommand{\ev}{\boldsymbol{e}}
\newcommand{\uv}{\boldsymbol{u}}
\newcommand{\KV}{\boldsymbol{K}}
\newcommand{\ro}{\boldsymbol{\rho}}
\newcommand{\si}{\boldsymbol{\sigma}}
\newcommand{\thv}{\boldsymbol{\theta}}
\newcommand{\bv}{\boldsymbol{b}}
\newcommand{\JV}{\boldsymbol{J}}
\newcommand{\nv}{\boldsymbol{n}}
\newcommand{\lv}{\boldsymbol{l}}
\newcommand{\om}{\boldsymbol{\omega}}
\newcommand{\Om}{\boldsymbol{\Omega}}
\newcommand{\Piv}{\boldsymbol{\Pi}}
\newcommand{\UV}{\boldsymbol{U}}
\newcommand{\iv}{\boldsymbol{i}}
\newcommand{\nuv}{\boldsymbol{\nu}}
\newcommand{\muv}{\boldsymbol{\mu}}
\newcommand{\lm}{\boldsymbol{\lambda}}
\newcommand{\Lm}{\boldsymbol{\Lambda}}
\newcommand{\opsi}{\overline{\psi}}
\renewcommand{\tan}{\textrm{tg}}
\renewcommand{\cot}{\textrm{ctg}}
\renewcommand{\sinh}{\textrm{sh}}
\renewcommand{\cosh}{\textrm{ch}}
\renewcommand{\tanh}{\textrm{th}}
\renewcommand{\coth}{\textrm{cth}}

\begin{abstract}
The properties of an anisotropic fluid outside a star or a black hole embedded in an expanding universe are investigated. One finds that, in Painleve-Gullstrand coordinates, the heat flux of the cosmological fluid vanishes, in spite of the nonzero value of a non-diagonal component of the stress tensor. The pressures and energy density of the fluid are regular at $r = 2m$, divergent at $r = 0$ and change sign at certain values of their arguments. 

 \end{abstract}

\section{Introduction}
 Isolated black holes have been investigated in great details in the last decades. However, black holes (BHs) embedded in an expanding universe are also important and even more realistic. The apparently simple question whether the cosmological expansion takes place even locally (at the level of an atom or of the Solar System) is a very complicate one and is still an unsolved problem \cite{CYS, KS}. Different authors gave different answers using different arguments \cite{WB}. The solution to the problem depends on the model of the universe (say, the commonly used Friedmann-Lemaitre-Robertson-Walker (FLRW)).
 
 The purpose is to combine both classes of solutions (local and cosmological) and to find exact solutions for the gravitational field of a compact object embedded in a cosmological background, namely a dynamical BH \cite{FJ, CG, SHMS, MA, FMN, FGCS}. The 1st author who has taken into consideration the effect of the cosmic expansion on local gravitational systems was McVittie \cite{MV}. He imposed a constraint on the (time dependent) mass of the central object for to avoid the accretion of the cosmic fluid into the central mass. The physical meaning of the McVittie solution of Einstein's equations is still under debate nowadays. A more recent metric describing a spherical mass in a cosmological background is the Sultana - Dyer space-time \cite{SD}. Their line element is conformal to the Schwarzschild one and the stress tensor corresponds to two noninteracting perfect fluids.
 
 Throughout the paper we use geometrical units $G = c = 1$, where $G$ is Newton's constant and $c$ is the velocity of light.
 
 \section{Painleve-Gullstrand coordinates}
 The Schwarzschild exact solution for the geometry outside a star or a BH
   \begin{equation}
  ds^{2} = -(1- \frac{2m}{r}) dt_{S}^{2} + (1- \frac{2m}{r})^{-1} dr^{2} + r^{2} d \Omega^{2}, 
 \label{2.1}
 \end{equation}
revealed a series of surprises: the incompleteness of the metric covered by the original Schwarzschild coordinates; a highly nontrivial global structure; the dynamic nature of the geometry, despite its static form revealed, for example, by Hawking's radiation \cite{KW}. In (2.1) $t_{S}$ is the Schwarzschild time and $d\Omega^{2}$ is the metric on the unit 2-sphere. To get rid of the apparent singularity of the metric at the horizon $r = 2m$, Painleve and Gullstarnd (P-G) used the following temporal transformation \cite{KW, TWZ, EP}
   \begin{equation}
   t = t_{S} + 2\sqrt{2mr}+ 2m ln\frac{\sqrt{r} - \sqrt{2m}}{\sqrt{r} + \sqrt{2m}} 
\label{2.2}
 \end{equation}
Therefore, the line element appears as 
   \begin{equation}
  ds^{2} = -(1- \frac{2m}{r}) dt^{2} + dr^{2} + 2\sqrt{\frac{2m}{r}} dt dr + r^{2} d \Omega^{2}. 
 \label{2.3}
 \end{equation}
 where $t$ is the free-fall time, that is the proper time experienced by an observer who free-falls from rest at infinity.
We chose the ''+'' sign in front of the square root in order to deal only with the inward moving free particles (along a geodesic curve with $dr + \sqrt{2m/r}dt = 0$, the velocity $dr/dt = - \sqrt{2m/r} $ is negative).

The geometry (2.3) is stationary, namely invariant under time translations (however, it is not invariant under time reversal because of the nondiagonal term). In addition, a constant time slice is simply flat space. We also emphasize that (2.3) represents physical space free falling radially into the BH at the Newtonian escape velocity $\sqrt{2m/r}$. The proper time of one observer at rest ($dr = d \theta = d \phi = 0$) is $d \tau = \sqrt{1 - (2m/r)}dt$. Moreover, the P-G metric paints a picture of space falling like a river (waterfall) into a spherical BH; its 4 - velocity evolves by a series of infinitesimal Lorentz boosts induced by the tidal changes in the river speed from place to place.

\section{Conformal Painleve-Gullstrand coordinates}
To look for the influence of the expanding universe on the local evolution of a BH, let us consider a time dependent P-G geometry given by
   \begin{equation}
  ds^{2} = a^{2}(t) \left[-(1- \frac{2m}{r}) dt^{2} + dr^{2} + 2\sqrt{\frac{2m}{r}} dt dr + r^{2} d \Omega^{2}\right], 
 \label{3.1}
 \end{equation}
 where the conformal factor $a(t)$ will be related to the scale factor in the FLRW universe. When $r >> 2m$ the metric (3.1) is conformally flat and therefore it corresponds to a spatially flat FLRW spacetime. 
 
 Our next task will be to impose the geometry (3.1) to be a solution of Einstein's equations
 \begin{equation}
 G_{ab} \equiv R_{ab} - \frac{1}{2} g_{ab}R^{c}_{~c} = 8 \pi T_{ab}   
 \label{3.2}
 \end{equation}
where the Latin indices run from $0$ to $3$. To reach that goal, the stress tensor $T_{ab}$ must have the nonzero components
\begin{equation}
\begin{split}
8 \pi T^{1}_{~1} =  \frac{ \dot{a}^{2}}{a^{4}} - \frac{2 \ddot{a}}{a^{3}} + \frac{4 \dot{a}}{a^{3}r} \sqrt{\frac{2m}{r}},~~8 \pi T^{0}_{~0} = - \frac{3 \dot{a}^{2}}{a^{4}} + \frac{3 \dot{a}}{a^{3}r} \sqrt{\frac{2m}{r}},\\
8 \pi T^{2}_{~2} = 8 \pi T^{3}_{~3} = \frac{\dot{a}^{2}}{a^{4}} - \frac{2 \ddot{a}}{a^{3}} + \frac{\dot{a}}{a^{3}r} \sqrt{\frac{2m}{r}},~~8 \pi T^{1}_{~0} = \frac{2(2 \dot{a}^{2} - a \ddot{a})}{a^{4}} \sqrt{\frac{2m}{r}} + \frac{2m \dot{a}}{a^{3}r^{2}}
\end{split}
\label{3.3}
\end{equation}
where $\dot{a}(t) = da/dt > 0$ and the indices $(0, 1, 2, 3)$ corresponds to $(t, r, \theta, \phi)$. The expressions (3.3) have been obtained thanks to the GrTensor II package and the software package Maple. The scalar curvature is given by   
 \begin{equation}
 R^{a}_{~a} = \frac{6 \ddot{a}}{a^{3}} - \frac{9 \dot{a}}{a^{3}r} \sqrt{\frac{2m}{r}} 
 \label{3.4}
 \end{equation}
 It is worth noting that $R^{a}_{~a}$ diverges when $r \rightarrow 0$, irrespective of the form of the function $a(t)$. However, the scalar curvature is regular at $r = 2m$ due to the regularity of the P-G metric there. That is in contrast with the results from \cite{FJ, CG, SHMS, HC1} where $R^{a}_{~a}$ is divergent not only at the origin $r = 0$ but also at the event horizon. We also observe that, far from the central mass (at $r >> 2m$) all components of $T^{a}_{~b}$ depend only on $a(t)$ and its time-derivative, as it should be. In addition, $T^{1}_{~1} = T^{2}_{~2} = T^{3}_{~3}$ and $T^{1}_{~0} = 0$, i.e. $T^{a}_{~b}$ is diagonal. 
 
 Let us take now a congruence of observers with the velocity vector field
  \begin{equation}
  u^{a} = \left(\frac{1}{a}, - \frac{1}{a} \sqrt{\frac{2m}{r}}, 0, 0\right) ,~~~u^{a}_{~a} = - 1.
 \label{3.5}
 \end{equation}
For an isolated BH we may put formally $a(t) = 1$ in (3.5) and $u^{1}$ becomes $- \sqrt{\frac{2m}{r}}$, that is the Newtonian escape velocity. It could be checked that with the ansatz (3.5) the energy-momentum tensor corresponds to an anisotropic fluid given by 
  \begin{equation}
  T_{ab} = (p_{\bot} + \rho) u_{a} u_{b} + p_{\bot} g_{ab} + (p_{r} - p_{\bot}) s_{a}s_{b} +  u_{a} q_{b} + u_{b} q_{a},
 \label{3.6}
 \end{equation}
with $\rho(r,t) = T_{ab}u^{a}u^{b} = -T^{0}_{~0}$ the energy density of the fluid, $p_{r}(r,t) = T^{1}_{~1}$ - the radial pressure, $p_{\bot} = T^{2}_{~2} = T^{3}_{~3}$ - the pressures on the transversal directions $\theta$ and $\phi$, $s^{a}$ is a spacelike vector orthogonal to $u^{a}$, with $s_{a}u^{a} = 0,~s_{a}s^{a} = 1$, $q^{a}$ is the heat flux with $q_{a}u^{a} = 0$ and it is given by the expression $q^{a} = - T^{a}_{~b}u^{b} - \rho u^{a}$, obtained from (3.6). Using now (3.3), (3.5) and (3.6) we find that 
  \begin{equation}
  s^{a} = \left(0, \frac{1}{a}, 0, 0 \right) ,~~~q^{a} = 0;~~~T^{a}_{~a} = p_{r} - \rho + 2p_{\bot}
 \label{3.7}
 \end{equation}
 In spite of the fact that $T^{1}_{~0} \neq 0$, we have obtained a vanishing heat flux $q^{a} = 0$. That is perhaps related to the geodesic character of the congruence (3.5). Indeed, it is easy to calculate that $a^{b} \equiv u^{a}\nabla_{a}u^{b} = 0$ (the observers of the congruence are in free fall).
 
 \section{Misner - Sharp energy}
The Misner-Sharp (MS) quasilocal energy $E(t,r)$ \cite{CG}, with its Weyl ($E_{W}$) and Ricci ($E_{R}$) parts, is useful to detect localized sources of gravity. In the case of spherical symmetry, $E$ is obtained from \cite{NY, CG2, HPFT} 
  \begin{equation}
  1 - \frac{2E(t,r)}{R} = g^{ab}R_{,a}R_{,b},
 \label{4.1}
 \end{equation}
where $R = a(t)r$ is the areal radius and $R_{,a} = \partial R/\partial x_{a}$. One finds, in the spacetime (3.1), that
  \begin{equation}
  E(t,r) = ma(t) + \frac{\dot{a}^{2}r^{3}}{2a} - \dot{a}r^{2}\sqrt{\frac{2m}{r}}
 \label{4.2}
 \end{equation}
 While the 1st term on the r.h.s. represents the Weyl energy $E_{W} = ma(t)$, the other two terms give the Ricci energy $E_{R}$. Keeping in mind that the cosmic fluid is anisotropic, $E_{R}$ may be written as \cite{BI, HC3} 
   \begin{equation}
     E_{R} = \frac{4 \pi}{3}R^{3}(\rho - p_{r} + p_{\bot}),
 \label{4.3}
 \end{equation}
 With $\rho,~p_{r}$ and $p_{\bot}$ from (3.3) we find that we indeed have
   \begin{equation}
  E(t,r) = \frac{\dot{a}^{2}r^{3}}{2a} - \dot{a}r^{2}\sqrt{\frac{2m}{r}}.
 \label{4.4}
 \end{equation}
 If the fluid were isotropic, we should have $p_{r} = p_{\bot}$ and, from (4.3), $ E_{R} = (\frac{4 \pi}{3})R^{3}\rho$, as expected.
 
 Let us further calculate the total radial energy crossing a $r = const.$ hypersurface $\Sigma$ \cite{GH} for some time interval 
    \begin{equation}
   W = \int T^{a}_{~b} u^{b} n_{a} \sqrt{- h}~ dt~ d\theta ~d\phi
 \label{4.5}
 \end{equation}
  where $n^{a} = (0, 1, 0, 0)$ is the unit spacelike normal vector on $\Sigma$, $h_{ab} = g_{ab} - n_{a}n_{b}$ is the induced metric on $\Sigma$ and $h = det(h_{ab})$. A direct evaluation of the integrand gives us $T^{a}_{~b} u^{b} n_{a} = 0$, where the equations (3.3) and (3.5) have been used. The fact that $W = 0$ is not surprising if we remember the P-G observers are in free fall (the acceleration vector $a^{b} = 0$) and the energy flux $q^{a}$ is vanishing.

 \section{Kinematical quantities and the Raychaudhuri equation}
 We have seen that the congruence (3.5) is geodesic so that the acceleration $a^{b}$ is zero. As far as the scalar expansion of the congruence is concerned, one obtains
   \begin{equation}
   \Theta \equiv \nabla_{a}u^{a} = \frac{3\dot{a}}{a^{2}} - \frac{3}{2ar}\sqrt{\frac{2m}{r}}.
 \label{5.1}
 \end{equation}
Keeping in mind that $a(t)$ and the radial coordinate are independent variables, $\Theta$ can be made positive or negative. In addition, it is nonzero when $a(t)$ = const., thanks to the fact that radial velocity is nonzero. The energy density and pressures from (3.3) show similar properties. This is an effect of the two competing factors: the stationary gravitational field of the central mass and the expansion of the universe. That is clear from (5.1) where the 2nd term on the r.h.s. is always negative and diminishes the 1st term. It is a consequence of the state of free fall of the ingoing fluid particles.

For the time evolution of the expansion, (5.1) yields
\begin{equation}
\dot{\Theta} \equiv u^{a}\nabla_{a}\Theta =  \frac{-6 \dot{a}^{2}}{a^{4}} + \frac{3\ddot{a}}{a^{3}} + \frac{3 \dot{a}}{2a^{3}r} \sqrt{\frac{2m}{r}} - \frac{9m}{2a^{2}r^{3}}.
\label{5.2}
\end{equation}
For an isolated BH ($a = 1$), we note that $\dot{\Theta} = -9m/2r^{3}$, even though the P-G metric is not time dependent ($\dot{\Theta} < 0$ is related to our choice for the radial velocity, that is $u^{1} < 0$). The shear tensor of the congruence \cite{HC2} is given by
\begin{equation}
\begin{split}
\sigma_{00} = \frac{2ma}{r^{2}}\sqrt{\frac{2m}{r}},~~~\sigma_{11} = \frac{a}{r}\sqrt{\frac{2m}{r}},~~~\sigma_{10} = \frac{2ma}{r^{2}} \\
\sigma_{22} = \frac{\sigma_{33}}{sin^{2}\theta} = -\frac{ar}{2}\sqrt{\frac{2m}{r}},~~~\sigma_{ab}\sigma^{ab} = \frac{3m}{a^{2}r^{3}}.
\end{split}
\label{5.3}
\end{equation}
Contrary to the previous results, the components (5.3) are of the form $f(a)~g(r)$ and, therefore, they vanish only when $r \rightarrow \infty$ (for $\sigma_{22}$ and $\sigma_{33}$, this is valid for the mixed components). They become, of course, null when $m = 0$, because of the homogeneity of the space. The Raychaudhuri equation
\begin{equation}
\dot{\Theta} + \frac{\Theta^{2}}{3} - \nabla_{b}a^{b} + \sigma_{ab}\sigma^{ab} - \omega_{ab}\omega^{ab} = - R_{ab}u^{a}u^{b}
\label{5.4}
\end{equation}
gives the time evolution of the scalar expansion in terms of the kinematical parameters and the Ricci tensor of the spacetime. The vorticity tensor $\omega_{ab}$ is zero for our congruence and also $a^{b}$. With $\Theta,~\dot{\Theta}$ and $\sigma_{ab}$ from, respectively, (5.1), (5.2) and (5.3), we see that (5.4) is satisfied, with $R_{ab}u^{a}u^{b}$ obtained from 
\begin{equation}
R_{ab}u^{a}u^{b} = 8 \pi (\rho + \frac{1}{2}T^{a}_{~a}).
\label{5.5}
\end{equation}
 We observe also that all the kinematical quantities are divergent at the singularity $r = 0$, irrespective of the value of $a(t)$.
 
 \section{de Sitter embedding of the P-G metric}
 Once we have analyzed the time dependent P-G space, we intend now to choose a particular expression of the scale factor $a(t)$ and study that case in detail. We shall follow the same procedure as in \cite{HC1}: take firstly $a(\bar{t}) = e^{H \bar{t}}$ and then change to $\eta = -(1/H) e^{-H \bar{t}}$ ($\bar{t} > 0,~-1/H < \eta <0$). We therefore embed the P-G black hole in a time dependent de Sitter spacetime and the metric is
 \begin{equation}
  ds^{2} = \frac{1}{H^{2}\eta^{2}} \left[-(1- \frac{2m}{r}) d \eta^{2} + dr^{2} + 2 \sqrt{\frac{2m}{r}} d \eta dr + r^{2} d \Omega^{2}\right], 
 \label{6.1}
 \end{equation}
with $a(\eta) = 1/H|\eta|$. For $r >> 2m$, (6.1) acquires the conformally flat de Sitter form. In contrast, with $a(\eta) = 1$ one obtains the stationary P-G line element. With the above special value for the scale factor $a(\eta)$, the energy density and pressures of the anisotropic fluid obtained from (3.3) are given by
\begin{equation}
\begin{split}
\rho = \frac{3H^{2}}{8 \pi} \left(1 - \frac{|\eta|}{r} \sqrt{\frac{2m}{r}}\right),~~p_{r} = \frac{3H^{2}}{8 \pi} \left(\frac{4|\eta|}{3r} \sqrt{\frac{2m}{r}} - 1\right),\\~~p_{\bot} = \frac{3H^{2}}{8 \pi} \left(\frac{|\eta|}{3r} \sqrt{\frac{2m}{r}} - 1 \right),
\end{split}
\label{6.2}
\end{equation}
with $\rho + p_{r} - p_{\bot} = 3H^{2}/8 \pi$, that may work as an equation of state. We note that asymptotically ($r >> 2m$) the fluid becomes isotropic ($p_{r} = p_{\bot}$) and $\rho = - p_{r}$, as expected for a de Sitter spacetime. The same effect is obtained if we put $\eta = 0$ in (6.2) (namely, $\bar{t} \rightarrow \infty)$. Moreover, $\rho, ~p_{r}$ and $~p_{\bot}$ remain finite at $r = 2m$. We also observe they change sign at certain value of $r$ and constant $\eta$ or at some value of $\eta$ and constant $r$. In other words, the weak energy condition is not satisfied for all ($\eta,r$) pairs.

Similar properties we find for the expansion $\Theta$ and the scalar curvature
\begin{equation}
\Theta = 3H \left(1 - \frac{|\eta|}{2r} \sqrt{\frac{2m}{r}}\right),~~~R^{a}_{~a} = 3H^{2} \left(4 - \frac{3|\eta|}{r} \sqrt{\frac{2m}{r}}\right). 
\label{6.3}
\end{equation}
While $\Theta$ and $R^{a}_{~a}$ become constant at $r >> 2m$ or $\eta \approx 0$, $8 \pi T^{1}_{~0} = 2mH^{2}|\eta|/r^{2}$ is negligible. One should be so for to recover the pure de Sitter values where $\Theta$ and $R^{a}_{~a}$ are indeed constant ($3H$, respectively $12H^{2}$) and the stress tensor is diagonal.

\section{Conclusions}
We investigated in this paper a compact object (a star or a black hole) in Painleve-Gullstrand coordinates and immersed in an expanding universe, paying special attention to the influence of cosmology on the local physics. The anisotropic fluid is freely falling at Newtonian escape velocity and the chosen congruence of observers is geodesic. Therefore, there is no heat flux despite the nonzero value of the $T^{1}_{~0}$ component of the energy momentum tensor. Excepting the shear tensor, the other parameters of the fluid (energy density, pressures, scalar expansion) change their sign at some value of $r$ at constant $\eta$, or of $\eta$ at constant $r$, a property that is difficult to have a reasonable physical interpretation (see, for example \cite{SHMS} where the authors reached similar conclusions).

\end{document}